\documentclass[sigconf]{acmart}

\usepackage{algorithm}
\usepackage{tikz}
\usepackage{algpseudocode}
\usepackage{amssymb}
\usepackage{amsmath}
\usepackage{algorithmicx}
\usepackage{algpseudocode}
\usepackage{comment}
\usepackage{color}
\usepackage{bm}
\usepackage{url}
\usepackage{enumitem}
\usepackage{booktabs}
\usepackage{multirow}
\usepackage{color,soul}
\usepackage{array}
\usepackage{tikz}
\usepackage{subcaption,graphicx}
\usetikzlibrary{fit,positioning}

\newcommand{\mt}{\mathsf{T}}
\DeclareMathOperator*{\argmax}{arg\,max}

\newcommand{\model}{{JNET}}

\begin{document}
\fancyhead{}


\copyrightyear{2020}
\acmYear{2020}
\setcopyright{acmcopyright}
\acmConference[WSDM '20]{The Thirteenth ACM International Conference on Web Search and Data Mining}{February 3--7, 2020}{Houston, TX, USA}
\acmBooktitle{The Thirteenth ACM International Conference on Web Search and Data Mining (WSDM '20), February 3--7, 2020, Houston, TX, USA}
\acmPrice{15.00}
\acmDOI{10.1145/3336191.3371770}
\acmISBN{978-1-4503-6822-3/20/02}

\title{JNET: Learning User Representations via Joint Network Embedding and Topic Embedding}
\author{Lin Gong, Lu Lin, Weihao Song, Hongning Wang}
\affiliation{
  \institution{Department of Computer Science, University of Virginia}
  \streetaddress{85 Engineer's Way, Charlottesville VA, 22904 USA}
}
\email{{lg5bt, ll5fy, ws5dw, hw5x} @virginia.edu}

\renewcommand{\shorttitle}{Joint Network Embedding and Topic Embedding}



\begin{abstract}

User representation learning is vital to capture diverse user preferences, while it is also challenging as user intents are latent and scattered among complex and different modalities of user-generated data, thus, not directly measurable. Inspired by the concept of user schema in social psychology, we take a new perspective to perform user representation learning by constructing a shared latent space to capture the dependency among different modalities of user-generated data. Both users and topics are embedded to the same space to encode users' social connections and text content, to facilitate joint modeling of different modalities, via a probabilistic generative framework. We evaluated the proposed solution on large collections of Yelp reviews and StackOverflow discussion posts, with their associated network structures. The proposed model outperformed several state-of-the-art topic modeling based user models with better predictive power in unseen documents, and state-of-the-art network embedding based user models with improved link prediction quality in unseen nodes. The learnt user representations are also proved to be useful in content recommendation, e.g., expert finding in StackOverflow.
\end{abstract}

%
%
%
%

\maketitle

\section{Introduction}

Inferring user intent from recorded user behavior data has been studied extensively for user modeling \cite{wang2011latent,shen2005implicit,liu2010personalized, gong2017clustered, wang2014user}. Essentially, user modeling builds up conceptual representations of users, which help automated systems to better capture users' needs and enhance user experience in such systems \cite{fischer2001user,kobsa2001generic}. The rapid development of social media enables users to participate in online activities and create vast amount of observational data, such as social interactions \cite{jin2001structure,kempe2003maximizing} and opinionated text content \cite{gong2018sentiment,pak2010twitter,deng2014exploring}, which in turn provides informative signs about user intents and enables more accurate user representation learning. Extensive efforts have proved the value of user representation learning in various real-world applications, such as latent factor models for collaborative filtering \cite{rendle2010factorization,koren2009matrix}, topic models for content modeling \cite{wang2011collaborative,mcauley2013hidden}, network embedding models for social link prediction \cite{liben2007link,bourigault2014learning}, and many more \cite{shen2005implicit,white2013enhancing}.

User representation learning is challenging, and it can never be a straightforward application of existing statistical learning algorithms on user-generated data. First, user-generated data is noisy, incomplete, highly unstructured, and tied with social interactions \cite{tang2014mining}, which imposes serious challenges in modeling such data. For example, in an environment where users are connected, e.g., social network, user-generated data is potentially related, which directly breaks the popularly imposed independent and identically distributed assumptions in most learning solutions \cite{liben2007link,tan2011user,gong2016modeling}. Second, users often participate in various online activities simultaneously, which creates instrumental contextual signals across different modalities.
Although oftentimes scattered and sparse, such multi-modal observations reflect users' underlying intents as a whole and call for a holistic modeling approach \cite{lecky1945self}. Ad-hoc data-driven solutions inevitably isolate the dependency and hence fail to create a comprehensive representation of users. For example, users' social interactions \cite{perozzi2014deepwalk,bourigault2014learning} and their generated text data \cite{blei2003latent,wang2011collaborative,mcauley2013hidden} have been extensively studied for user representation learning, but they are largely modeled in isolation. Third, consequently, a unified user representation learning solution is preferred to serve different applications, by taking advantage of data-rich applications to help those data-poor applications. 

Even among a few attempts for joint modeling of different types of user-generated data \cite{yang2015network,gong2018sentiment}, \textit{explicit modeling of dependency} among multiple behavior modalities is still missing. For example, Yang et al. \cite{yang2015network} incorporated user-generated text content into network representation learning via joint matrix factorization. In their solution, content modeling is only used as a regularization for network modeling; and thus the learnt model is not in a position to predict unseen text content. Gong and Wang \cite{gong2018sentiment} paired the task of sentiment classification with that of social network modeling, and represented each user as a mixture over the instances of these paired tasks. Though text and network are jointly considered, they are only correlated by sharing the same mixing component, without explicitly modeling of the mutual influence between them. 

In social psychology and cognitive science, the concept of \textit{user schema} defines the knowledge structure a person holds which organizes categories of information and the relationships among such categories \cite{tuckey2003influence}. Putting it into the scenario of user modeling, we naturally interpret the knowledge structure as user representation described by the collection of associated data, such as the set of textual reviews and behavioral logs associated with individual users. The interrelation existing among multiple types of data further motivates us to perform user modeling in a joint manner while the concept of distributed representation learning \cite{bengio2013representation}, i.e., embedding, provides us one possible solution. By constructing a shared latent space, we can embed different modalities of user-generated data in the same low-dimensional space, where the structural dependency among them can be realized by the proximity among different embeddings. The space should be constructed in such a way: 1) the properties of each modality of user-generated data is preserved; 2) the closeness among different modalities of user-generated data can be characterized by the similarity measured in the latent space. For example, connected users in a social network should be closer to each other in this latent space; and by mapping other types of user behavior data into this same space, e.g., text data or behavioral logs, users should be surrounded by their own generated data.


To realize this new perspective of user representation learning, we exploit two most widely available and representative forms of user-generated data, i.e., text content and social interactions. We develop a probabilistic generative model to integrate user modeling with content and network embedding. Due to the unstructured nature of text, we appeal to statistical topic models to model user-generated text content \cite{blei2003latent, wang2011collaborative}, with a goal to capture the underlying semantics. We define a topic as a probability distribution over a fixed vocabulary \cite{blei2003latent}. We embed both users and topics to the same low-dimensional space to capture of their mutual dependency. On one hand, a user's affinity to a topic is characterized by his/her proximity to the topic's embedding in this latent space, which is utilized to generate each text document of the user. On the other hand, the affinity between users is directly modeled by the proximity between users' embeddings, which are utilized to generate the corresponding social network connections. In this latent space, the two modalities of user-generated data are correlated explicitly, indicated by the user's topical preferences. The user representation is obtained by posterior inference of those embedding vectors over a set of training data, via variational Bayesian. To reflect the nature of our proposed user representation learning method, we name the solution \textbf{J}oint \textbf{N}etwork \textbf{E}mbedding and \textbf{T}opic Embedding, or \model{} for short. 

Extensive empirical evaluations are performed on two large collections of user-generated text documents from Yelp and StackOverflow, together with their network structures. Compared with a set of state-of-the-art user representation learning solutions, clear advantages of \model{} are observed: the model's predictive power in content modeling is enhanced on users with rich social connections, and similar improvement is observed in its prediction in network modeling on users with rich text data. The use of learnt user representation generalizes beyond content modeling and social network modeling: it accurately suggests technical discussion threads for users to participate in StackOverflow, e.g., expert recommendation.

\section{Related Work}
In order to learn effective user representations, a lot efforts have been devoted to modeling diverse modalities of user-generated data: 1) in an isolated manner, i.e., focusing on one particular modality of user-generated data such as text reviews; 2) in a joint manner, i.e., utilizing multiple types of user data. Our proposed solution falls into the second category as it learns user representations from both network structure and text content by  explicitly capturing the dependency between the two modalities in the latent topic space. 


When performing user representation learning in an isolated way, much attention has been paid on exploring user-user interactions to learn users' distributed representations, which are essential for better understanding users' interactive preferences in social network analysis. Inspired from word embedding techniques \cite{mikolov2013efficient}, random walk models are exploited to generate random paths over a network to learn dense, continuous and low-dimensional representations of users \cite{perozzi2014deepwalk, tang2015line, grover2016node2vec}. 
Matrix factorization technique is also commonly used to learn user embeddings \cite{ou2016asymmetric, wang2017community}, as learning a low-rank space for an adjacency matrix representing the network naturally fits the need of learning low-rank user/node embeddings. For instance, Tang and Liu \cite{tang2009relational} factorize an input network's modularity matrix and use discriminative training to extract representative dimensions for learning user representation. 

In parallel, the user-generated text data is utilized to understand users' emphasis on specific entities or aspects. Topic models \cite{blei2003latent, hofmann1999probabilistic} serve as a building block for statistical modeling of text data. Typical solutions model individual users as a bag of topics \cite{rosen2004author}, which govern the generation of associated text documents. Wang and Blei \cite{wang2011collaborative} combine topic modeling with collaborative filtering to estimate topical user representations with additional observations from user-item ratings. Wang et al. \cite{wang2011latent} use topic modeling to estimate users' detailed aspect-level preferences from their opinionated review content. Lin et al. \cite{lin2019learning} learn users' personalized topical compositions to differentiate user's subjectivity from item's intrinsic property in the review documents. McAuley and Leskovec \cite{mcauley2013hidden} uncover the implicit preferences of each user as well as the properties of each product by mapping users and items into a shared topic space. Some recent works use deep neural networks to obtain user embedding from their generated text data \cite{tang2015learning, chen2016neural}.

Although most previous works studied social networks and user-generated text content in isolation, little attention has been paid in combining the two sources for better user modeling. Earlier work \cite{mei2008topic} regularizes a statistical topic model with a harmonic regularizer defined on the network structure. Yang et al. \cite{yang2015network} incorporate text features of users into network representation learning via joint matrix factorization. Gong and Wang \cite{gong2018sentiment} pair tasks of opinionated content modeling and network structure modeling in a group-wise fashion, and model each user as a mixture over the tasks. Though both text and network are utilized for user modeling in the aforementioned works, explicit modeling of dependence among different modalities is still missing. Archarya et al. \cite{acharya2015gamma} explore the dependency among documents and network but on a per-community basis instead of a per-user basis. Our work proposes a holistic view to model users' social preferences and topical interests jointly, thus to provide a more general understanding of user intents from multiple perspectives. 


\section{Joint Network Embedding and Topic Embedding}

To interrelate different modalities of user-generated data for user representation learning, we propose to perform joint network embedding and topic embedding. In this section, we first provide the details of our probabilistic generative model, \model{}, which imposes a complete generative process over user-generated social interactions and text data in each individual user. Then we describe our variational Bayesian based Expectation Maximization algorithm, which retrieves the learnt user representation from a given corpus. 

\subsection{Model Specification}
We denote a collection of $U$ users as $\mathcal{U}=\{u_1, u_2, ...u_{U}\}$, in which each user $u_i$ is associated with a set of documents $\mathcal{D}_i\!=\!\big\{ x_{i,d}\big\}^{D_i}_{d=1}$. Each document is represented as a bag of words $x_d=\{w_1, w_2,..,w_N\}$, where each $w_n$ is chosen from a vocabulary of size $V$. Each user is also associated with a set of social connections denoted as $\mathcal{E}_i=\{e_{ij}\}_{j\neq i}^U$, where $e_{ij}=1$ indicates user $u_{i}$ and $u_j$ are connected in the network; otherwise, $e_{ij}=0$.



We represent each user as a real-valued continuous vector $u_i\in \mathbb{R}^M$ in a low-dimensional space. And we seek to impose a joint distribution over the observations in each user's associated text documents and social interactions, so as to capture the underlying structural dependency between these two types of data. Based on our assumption that both types of users-generated data are governed by the same underlying user intent, we explicitly model the joint distribution as $p(\mathcal{D}_i, \mathcal{E}_i)=\int p(\mathcal{D}_i, \mathcal{E}_i, u_i) \mathrm{d}u_i$, which can be further decomposed into $p(\mathcal{D}_i, \mathcal{E}_i, u_i)=p(\mathcal{D}_i| \mathcal{E}_i, u_i)p(\mathcal{E}_i| u_i)p(u_i)$. We assume given the user representation $u_i$, the generation of text documents in $\mathcal{D}_i$ is independent from the generation of social interactions in $\mathcal{E}_i$, i.e., $p(\mathcal{D}_i| \mathcal{E}_i, u_i)=p(\mathcal{D}_i|u_i)$. As a result, the modeling of joint probability over a user's observational data with his/her latent representation can be decomposed into three related modeling tasks: 1) $p(\mathcal{D}_i|u_i)$ for content modeling, 2) $p(\mathcal{E}_i|u_i)$ for social connection modeling, and 3) $p(u_i)$ for user embedding modeling. 

We appeal to topic models \cite{blei2003latent, hofmann1999probabilistic} due to their effectiveness shown in existing empirical studies for content modeling. The concept of user schema inspires us to embed both users and topics to the same latent space in order to capture the dependency between them. By projecting a user's embedding vector to topic embedding vectors, we can easily measure affinity between a user and a topic, and thus capture users' topical preferences. It also allows us to capture the topical variance in documents from the same user and establish a valid predictive distribution of his/her documents. 


Formally, we assume there are in total $K$ topics underlying the corpus with each represented as an embedding vector $\phi_k \in \mathbb{R}^M$ in the same latent space; denote $\Phi \in \mathbb{R}^{K\times M}$ as the matrix of topic embeddings, which facilitate our representation of each user's affinity towards different topics: $\Phi\cdot u_i$, 
which reflects user $u_i$'s topical preferences, and serves as the prior of topic distribution in each text document from him/her. Specifically, denote the document-level topic vector as $\theta_{id}\in \mathbb{R}^K$, we have $\theta_{id} \sim \mathcal{N}(\Phi\cdot u_i, \tau^{-1}{I})$, where $\tau$ characterizes the uncertainty when user $u_i$ is choosing topics from his/her global topic preferences for each single document. 
By projecting the document-level topic vector $\theta_{id}$ into a probability simplex, we obtain the topic distribution for document $x_{i,d}$:  $\pi_{idk}=\text{softmax}(\theta_{idk})=\exp(\theta_{idk})/\sum_{l=1}^K\exp(\theta_{idl})$, from which we sample a topic indicator $z_{idn}\in\{1, ..., K\}$ for each word $w_{idn}$ in $x_{i,d}$ by $z_{idn} \sim \text{Multi}(\pi_{idk})$. As in conventional topic models, each topic $k$ is also associated with a multinomial distribution $\beta_k$ over a fixed vocabulary, and each word $w_{idn}$ is then drawn from the respective word distribution indicated by corresponding topic assignment, i.e., $w_{idn}\sim p(w|\beta_{z_{idn}})$. Putting all pieces together, the task of content modeling for each user can be summarized as $p(\mathcal{D}_i|u_i)=\prod_{d=1}^{D_i}p(\theta_{id}|u_i, \Phi, \tau)\prod_{n=1}^Np(z_{idn}|\theta_{id})p(w_{idn}|z_{idn},\beta)$.

The key in modeling social connections is to understand the closeness among users. As we represent users with a real-valued continuous vector, this can be easily measured by the vector inner product in the learnt low-dimensional space. Define the underlying affinity between a pair of users $u_i$ and $u_j$ as $\delta_{ij}$, we assume $\mathbb{E}[\delta_{ij}]=u_i^\mt u_j$.
To capture uncertainty of the affinity between different pairs of users, we further assume $\delta_{ij}$ is drawn from a Gaussian distribution centered at the measured closeness, $\delta_{ij}\sim \mathcal{N}(u_i^\mt u_j, \xi^2)$, where $\xi$ characterizes the concentration of this distribution. 
The observed social connection $e_{ij}$ between user $u_i$ and $u_j$ is then assumed as a realization of this underlying user affinity: $e_{ij} \sim \text{Bernoulli}(\text{logistic}(\delta_{ij}))$ where $\text{logistic}(\delta_{ij})=1/(1+\exp(-\delta_{ij}))$. As a result, the task of social connection modeling can be achieved by $p(\mathcal{E}_i|u_i)=\prod_{j\neq i}^Up(e_{ij}|\delta_{ij})p(\delta_{ij}|u_i, u_j)$.

\begin{figure}[t]
\centering
\vspace{-5mm}
\setlength\tabcolsep{.5pt}
  \includegraphics[width=0.98\linewidth]{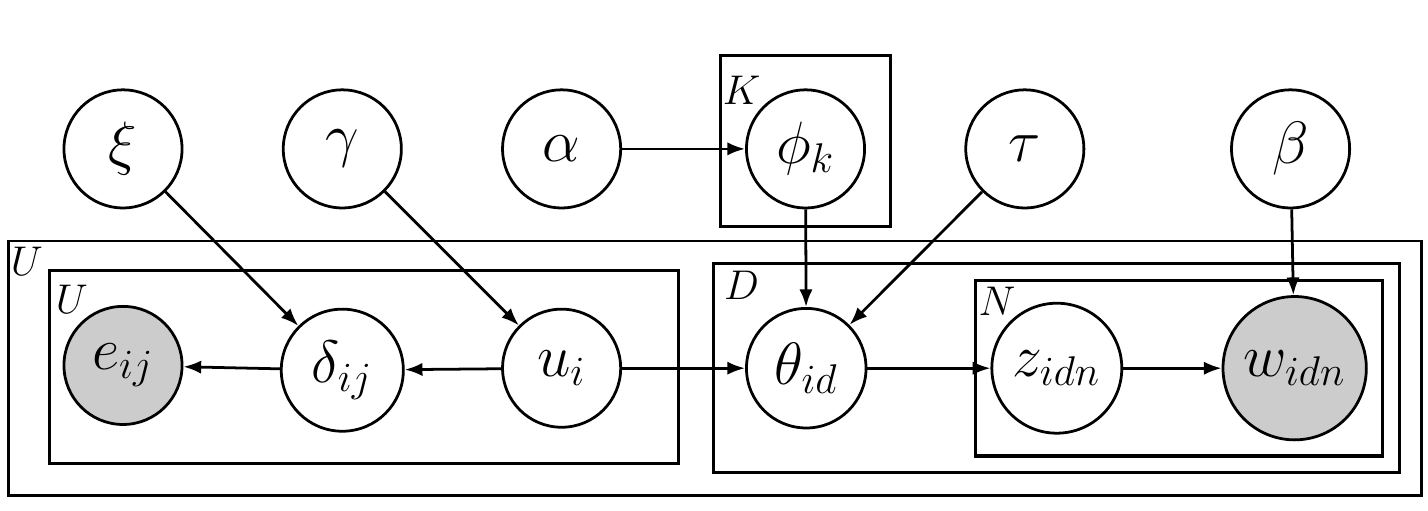}
\caption{Graphical model representation of \model{}. The upper plate indexed by $K$ denotes the learnt topic embeddings. The outer plate indexed by $U$ denotes distinct users in the collection. The inner plates indexed by $U$ and $D$ denote each user's social connections and text documents respectively. The inner plate indexed by $N$ denotes the word content in one text document.}
\label{fig_model}
\vspace{-8mm}
\end{figure}

We do not have any specific constraint on the form of latent user embedding vectors $\{u_i\}^U_{i=1}$ and topic embedding $\{\phi_{k}\}^{K}_{k=1}$, as long as they are in a $M$-dimensional space. For simplicity, we assume they are drawn from isotropic Gaussian distributions respectively, i.e., $u_i\sim \mathcal{N}(\bm{0},\gamma^{-1}\bm{I})$, where $\gamma$ measures the concentration of different users' embedding vectors, and $\phi_k \sim \mathcal{N}(\bm{0},\alpha^{-1}\bm{I})$. Other types of prior distribution can also be introduced, if one has more knowledge about the user and topic embeddings, such as sparsity or a particular geometric shape. But it is generally preferred to have conjugate priors, so as to simplify later posterior inference steps.

Putting these components together, the generative process of our solution can be described as follows:
\begin{itemize}
\item For each topic $\phi_k$:
    \begin{itemize}
    \item Draw its topic compact representation $\phi_k \sim \mathcal{N}(\bm{0},\alpha^{-1}\bm{I})$
    \end{itemize}
\item For each user $u_i$:
    \begin{itemize}
	\item Draw its user compact representation $u_i\sim \mathcal{N}(\bm{0},\gamma^{-1}\bm{I})$
	\item For every other user $u_j$:
        \begin{itemize}
		\item Draw affinity $\delta_{ij}$ between $u_i$ and $u_j$, $\delta_{ij} \sim \mathcal{N}(u_i^\mt u_j, \xi^2)$
		\item Draw interaction $e_{ij}$ between $u_i$ and $u_j$, $e_{ij} \sim \\  \text{Bernoulli}(\text{logistic}(\delta_{ij}))$
		\end{itemize}
	\end{itemize}
\item For each document of user $u_i$:
    \begin{itemize}
    \item Draw the user-document topic preference vector \\$\theta_{id}\sim \mathcal{N}(\Phi\cdot u_i ,\tau^{-1}\bm{I})$
    \item For each word $w_{idn}$:
        \begin{itemize}
   	    \item Draw topic assignment $z_{idn}\sim \text{Multi(softmax}(\theta_{id}))$
    	\item Draw word $w_{idn}\sim \text{Multi}(\beta_{z_{idn}})$
    	\end{itemize}
    \end{itemize}
\end{itemize}


We make two explicit assumptions here: 1) the dimensionality $M$ of the compact representation of topics and users is predefined and fixed; 2) the word distributions under topics are parameterized by a $K\times V$ matrix $\beta$ where $\beta_{kv}=p(w^v|z^k)$ over a fixed vocabulary of size $V$. The generative model captures the interrelation between multiple modalities of user-generated data for user representation learning. In essence, we are performing a \textbf{J}oint \textbf{N}etwork \textbf{E}mbedding and \textbf{T}opic Embedding, thus, we name the resulting model as \model{} in short.

\subsection{Variational Bayesian Inference}
\label{sec_infer}
The compact user representations can be obtained via posterior inference over the latent variables on a given set of data. However, posterior inference is not analytically tractable in \model{} due to the coupling among latent variables, i.e., user-user affinity ${\delta}$, user embedding ${u}$, topic embedding $\Phi$, document-level topic proportion $\theta$ and word-level topic assignment ${z}$. We appeal to a mean-field variational method to  approximate the posterior distributions, and further utilize Taylor expansion \cite{blei2006correlated} to address the difficulty introduced by non-conjugate logistic-normal priors.

We begin by postulating a factorized distribution:
$q(\Phi, U, \Delta, \Theta, Z)=$ $\prod_{k=1}^Kq(\phi_k) \prod_{i=1}^Uq(u_i) \Big[\prod_{j=1,j\neq i}^Uq(\delta_{ij})\prod_{d=1}^Dq(\theta_{id})\prod_{n=1}^Nq(z_{idn})\Big]$,
where the factors have the following parametric forms:
\begin{align}
&q(\phi_k)=\mathcal{N}(\phi_k|\mu^{(\phi_k)},\Sigma^{(\phi_k)}), q(u_i)=\mathcal{N}(u_i|\mu^{(u_i)},\nonumber
\Sigma^{(u_i)}),\nonumber\\&q(\delta_{ij})=\mathcal{N}(\delta_{ij}|\mu^{(\delta_{ij})}, {\sigma^{(\delta_{ij})}}^2),\nonumber
q(\theta_{id})=\mathcal{N}(\theta_{id}|\mu^{(\theta_{id})},\Sigma^{(\theta_{id})}),\\&q(z_{idn})=\text{Mult}(z_{idn}|\eta_{idn})\nonumber
\end{align}
Because the topic proportion vector $\theta_{id}$ is inferred in each document, it is not necessary to estimate a full covariance matrix for it \cite{blei2006correlated}. Hence, in its variational distribution, we only estimate the diagonal variance parameters. 

Variational algorithms aim to minimize the KL divergence from the approximated posterior distribution $q$ to the true posterior distribution $p$. It is equivalent to tightening the evidence lower bound (ELBO) by Jensen's inequality \cite{blei2003latent}:
\begin{align}
\label{eq_elbo}
&\log p(\bm w, \bm e |\alpha, \beta, \gamma, \tau)
\\
& \geq \mathbb{E}_{q}[\log p(U, \Theta, Z, \Phi, \Delta, \bm w, \bm e|\alpha, \beta, \gamma, \tau)] -\mathbb{E}_{q}[\log q(U, \Theta, Z, \Phi, \Delta)] \nonumber
\end{align}
where the expectation is taken with respect to the factorized variaitonal distribution of the latent variables $q(\Phi, U, \Delta, \Theta, Z)$. Let $\mathcal{L}(q)$
denote the right-hand side of Eq \eqref{eq_elbo}, the first step of maximizing this lower bound is to derive the analytic form of posterior expectations required in $\mathcal{L}(q)$. Thanks to the conjugate priors introduced on $\{{u_i}\}_{i=1}^U$ and $\Phi=\{\phi_k\}_{k=1}^K$, the expectations related to these latent variables have closed form solutions, while due to non-conjugate logistic-normal priors, we use Taylor expansions to approximate the expectations related to $\theta_{id}, 
\delta_{ij}$. Next we describe the detailed inference procedure for each latent variable.

\noindent\textbf{$\bullet$ Estimate topic embedding.} For each topic $k$, we relate the terms associated with $q(\phi_k|\mu^{(\phi_k)},\Sigma^{(\phi_k)})$ in Eq \eqref{eq_elbo} and take maximization w.r.t. $\mu^{(\phi_k)}$ and $\Sigma^{(\phi_k)}$. Closed form estimations of $\mu^{(\phi_k)},\Sigma^{(\phi_k)}$ exist,
\begin{equation}
\begin{split}
&\mu^{(\phi_k)}=\tau\Sigma^{(\phi_k)}\sum\nolimits_{i=1}^U\sum\nolimits_{d=1}^{D_i}\mu_k^{(\theta_{id})}\mu^{(u_i)} 
\\&\Sigma^{(\phi_k)}=\big[\alpha \bm I+\tau\sum\nolimits_{i=1}^U\sum\nolimits_{d=1}^{D_i}(\Sigma^{(u_i)}+\mu^{(u_i)} {\mu^{(u_i)}}^\mt)\big]^{-1} 
\label{sigma_phi_k}
\end{split}
\end{equation}

The estimation of $\Sigma^{(\phi_k)}$ is not related to a specific topic $k$, because we impose an isotropic Gaussian prior for all $\{\phi_k\}^K_{k=1}$ in \model{}. It suggests that the correlations between different topic embedding dimensions are homogeneous across topics. 
Interestingly, we can notice that the posterior covariance $\Sigma^{(\phi_k)}$ of topic embeddings is closely related to user embeddings, which indicates direct dependency from network structure to text content. 

\noindent\textbf{$\bullet$ Estimate user embedding.} For each user $i$, we relate the terms associated with $q(u_i|\mu^{(u_i)},\Sigma^{(u_i)})$ in Eq \eqref{eq_elbo} and maximize it with respect to $\mu^{(u_i)},\Sigma^{(u_i)}$. Closed form estimations can also be achieved for these two parameters as follows:
\begin{align}
\mu^{(u_i)}=&\Sigma^{(u_i)}\big(\tau\sum\nolimits_{d=1}^{D_i}\sum\nolimits_{k=1}^K\mu_k^{(\theta_{id})}\mu^{(\phi_k)}+\sum\nolimits_{j\neq i}^U\xi^{-2}\mu^{(\delta_{ij})}\mu^{(u_j)}\big)\nonumber\\
\Sigma^{(u_i)}=&\gamma \bm I+\tau D_i\sum\nolimits_{k=1}^K(\Sigma^{(\phi_k)}+\mu^{(\phi_k)}{\mu^{(\phi_k)}}^\mt) \nonumber\\
&+\sum\nolimits_{j\neq i}^U\xi^{-2}(\Sigma^{(u_j)}+\mu^{(u_j)}{\mu^{(u_j)}}^\mt) 
\label{posterior_user}
\end{align}

The effect of joint content modeling and network modeling for user representation learning is clearly depicted in this posterior estimation of user embedding vectors. The updates of $\mu^{(u_i)}$ and $\Sigma^{(u_i)}$ come from two types of influence: the text content and social interactions of the current user. For example, the posterior mode estimation of user embedding vector $u_i$ is a weighted average over the topic vectors that this user has used in his/her past text documents and the user vectors from his/her friends. And the weights measure his/her affinity to those topics and users in each specific observation. The updates exactly reflect the formation of ``user schema'' in social psychology from two perspectives: both modalities of user-generated data shape user embeddings, while the structural dependency between them is reflected in this unified user representation.

\noindent\textbf{$\bullet$ Estimate per-document topic proportion vector.} Similar procedures as above can be taken to estimate $\mu^{(\theta_{id})}$ and $\Sigma^{(\theta_{id})}$. Due to the lack of conjugate prior for logistic Normal distributions, we apply Taylor expansion and introduce an additional free variational parameter $\zeta$ in each document. Because there is no closed form solution for the resulting optimization problem, we use gradient ascent to optimize $\mu^{(\theta_{id})}$ and $\Sigma^{(\theta_{id})}$ with the following gradients,
\begin{equation}
\begin{split}
\label{eq_theta}
\partial L / \partial \mu_{k}^{(\theta_{id})}=&-\tau\mu_{k}^{(\theta_{id})}+\tau {\mu^{(\phi_k)}}^\mt\mu^{(u_i)}
\\
&+\sum\nolimits_{n=1}^N\big[\eta_{idnk}-\zeta^{-1}\exp(\mu_k^{(\theta_{id})}+\Sigma_{kk}^{(\theta_{id})}/2)\big]\\
\partial L / \partial \Sigma_{kk}^{(\theta_{id})}=&-\tau-N\exp(\mu_k^{(\theta_{id})}+\Sigma_{kk}^{(\theta_{id})}/2)/\zeta+1/\Sigma_{kk}^{(\theta_{id})}
\end{split}
\end{equation}
where $\zeta=\sum_{k=1}^K\exp(\mu_k^{(\theta_{id})}+\Sigma_{kk}^{(\theta_{id})}/2)$. Since only the diagonal elements in $\Sigma^{(\theta)}_{id}$ are statistically meaningful (i.e., variance), we simply set its off-diagonal elements to zero in gradient update. 
The gradient function suggests that the document-level topic proportion vector should align with the corresponding compact user representation and topic representation. 
Although no closed form estimations of $\mu^{(\theta_{id})}$ and $\Sigma^{(\theta_{id})}$ exist, the expected property of $\mu^{(\theta_{id})}$ is clearly reflected: the proportion of each topic in document $x_{i,d}$ should align with this user's preference on this topic (i.e., affinity in the embedding space) and the topic assignment in document content. And the variance is introduced by the uncertainty of per-word topic choice and the intrinsic uncertainty of a user's affinity with a topic.  

\noindent\textbf{$\bullet$ Estimate user affinity.} Similar approach can be applied here to estimate $\mu^{(\delta_{ij})}$ and ${\sigma^{(\delta_{ij})}}^2$ which govern the latent user affinity. Again, gradient ascent is utilized to optimize $\mu^{(\delta_{ij})}$ and $\Sigma^{(\theta_{id})}$,
\begin{align*}
\!\!&\partial L / \partial \mu^{(\delta_{ij})}=e_{ij}-\varepsilon^{-1}\exp{(\mu^{(\delta_{ij})}+{\sigma^{(\delta_{ij})}}^2/2)}-\xi^{-2}(\mu^{(\delta_{ij})}-{\mu^{(u_i)}}^\mt\mu^{(u_j)})
\\
\!\!&\partial L/\partial  \sigma^{(\delta_{ij})}=-\varepsilon^{-1}\sigma^{(\delta_{ij})}\exp{(\mu^{(\delta_{ij})}+{\sigma^{(\delta_{ij})}}^2/2)}-\xi^{-2}\sigma^{(\delta_{ij})}+1/\sigma^{(\delta_{ij})}
\end{align*}
The gradient functions suggest that the latent affinity between a pair of users is closely related with their observed connectivity and their closeness in the embedding space. 

\noindent\textbf{$\bullet$ Estimate word topic assignment.} The topic assignment $z_{idn}$ for each word $w_{idn}$ in document $x_{i,d}$ can be estimated by,
\begin{equation*}
\eta_{idnk}\propto \exp\{\mu_k^{(\theta_{id})}+\sum\nolimits_{v=1}^Vw_{idnv}\log \beta_{kv}\}
\end{equation*}

We execute the above variational inference procedures in an alternative fashion until the lower bound $\mathcal{L}(q)$ defined in Eq \eqref{eq_elbo} converges. The variational inference algorithm postulates strong independence structures between the variational parameters, allowing straightforward \textbf{parallel computing}. Since the variational parameters can be grouped by documents: $\mu^{(\theta_{id})}$, $\Sigma^{(\theta_{id})}$ and $\eta$, by topics: $\mu^{(\phi_k)}$ and $\Sigma^{(\phi_k)}$, and by users: $\mu^{(u_i)}$, $\Sigma^{(u_i)}$, $\mu^{(\delta_{ij})}$ and ${\sigma^{(\delta_{ij})}}^2$, we perform alternative update in parallel to improve computational efficiency: for example, we fix topic-level parameters and user-level parameters, and distribute the documents across different machines to estimate their own $\mu^{(\theta_{id})}$, $\Sigma^{(\theta_{id})}$ and $\eta$ in parallel for large collections of user-generated data.  

\subsection{Parameter Estimation}

When performing the variational inference described above, we have assumed the knowledge of model parameters $\alpha, \gamma, \tau, \xi$ and $\beta$. Based on the inferred posterior distribution of latent variables in \model{}, these model parameters can be readily estimated by the Expectation-Maximization (EM) algorithm. The most important model parameters are priors for user embedding $\gamma$ and topic embedding $\alpha$, and word-topic distribution $\beta$. As $\xi$ and $\tau$ serve as the variance for user affinity $\delta_{ij}$ and document topic proportion vector $\theta_{id}$, and we have large amount of observations in text documents and social connections across all users, our model is less sensitive to their settings. Therefore, we estimate $\xi$ and $\tau$ less frequently than $\alpha$, $\gamma$ and $\beta$.

By taking the gradient of $\mathcal{L}(q)$ in Eq \eqref{eq_elbo} with respect to $\alpha$, and set the resulting gradient to 0, we get the closed form estimation of $\alpha$ as follows:
\begin{align*}
\alpha = \frac{KM}{\sum_{k=1}^K[\sum_{m=1}^M\Sigma_{mm}^{(\phi_k)}+{\mu^{(\phi_k)}}^\mt \mu^{(\phi_k)}]},
\end{align*}
Similarly, the closed form estimation of $\gamma$ can be easily derived as,
\begin{align*}
\gamma = \frac{UM}{\sum_{i=1}^U[\sum_{m=1}^M\Sigma_{mm}^{(u_i)}+{\mu^{(u_i)}}^\mt \mu^{(u_i)}]}.
\end{align*}
And the closed form estimation for word-topic distribution $\beta$ can be achieved by,
\begin{align*}
\beta_{kv} \propto \sum\nolimits_{i=1}^U\sum\nolimits_{d=1}^{D_i}\sum\nolimits_{n=1}^{N}w_{idnv} \eta_{idnv},
\end{align*}
where $w_{idnv}$ indicates the $n$th word in $u_i$'s $d$th document is the $v$th term in the vocabulary. The estimation for $\xi$ and $\tau$ is omitted for space limit, but they can be easily derived based on Eq \eqref{eq_elbo}.


The resulting EM algorithm consists of E-step and M-step. In E-step, the variational parameters are inferred based the procedures described in Section \ref{sec_infer}; and in M-step, the model parameters are estimated based on collected sufficient statistics from E-step. These two steps are repeated until the lower bound $\mathcal{L}(q)$ converges over all training data.


Inferring the latent variables with each user and each topic are computationally cheap. Specifically, by Eq~\eqref{sigma_phi_k}, updating the variables for each topic imposes a complexity of $\mathcal{O}\big(KM^2|D|\big)$, where $K$ is the total number of topics, $M$ is the latent dimension, $|D|$ is the total number of documents. By Eq~\eqref{posterior_user}, updating the variables for each user imposes a complexity of $\mathcal{O}(M^2U^2)$ where $U$ is the total number of users. Estimating the latent variables for the per-document topic proportion imposes a complexity of $\mathcal{O}(|D|K(\bar N + M))$ by Eq~\eqref{eq_theta}, where $\bar N$ is the average document length. And updating variables for each pair of user affinity takes constant time while there are $U^2$ affinity variables. With the consideration of the total number of users and topics, the overall complexity for the proposed algorithm is $\mathcal{O}(KM^2|D| + M^2U^2)$.



\section{Experiments}
\begin{figure*}[t] 
\centering
\includegraphics[width=17.2cm]{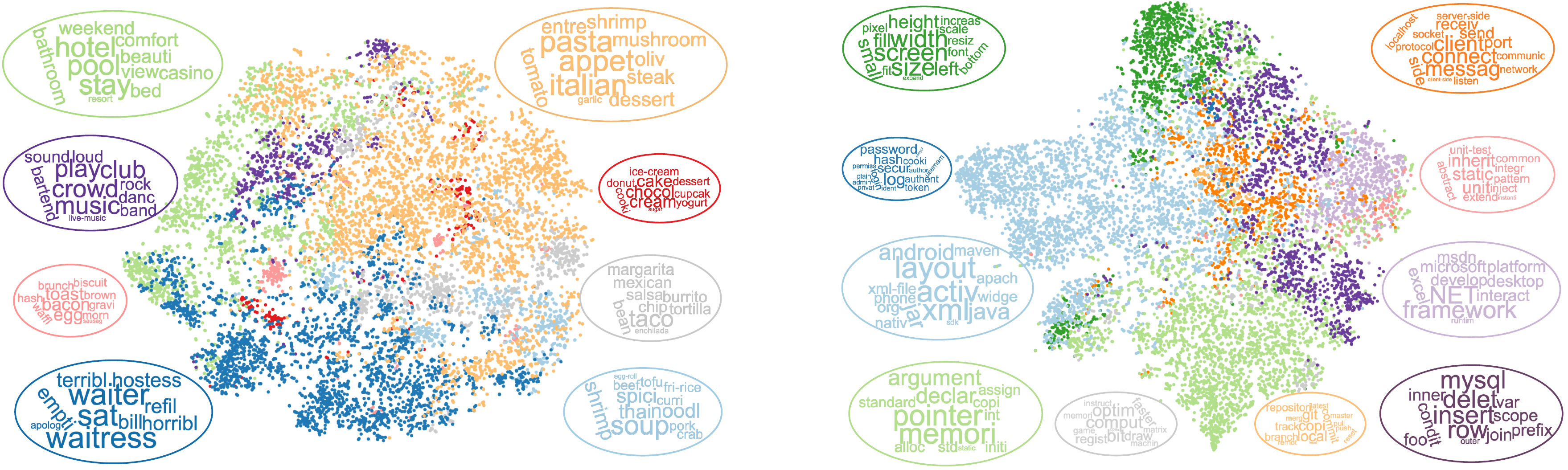} 
\vspace{-3mm}
\caption{Visualization of user embedding and learnt topics in 2-D space of Yelp (left) and StackOverflow (right).}
\label{frame}
\vspace{-3mm}
\end{figure*} 

We evaluated the proposed model on large collections of Yelp reviews and StackOverflow forum discussions, together with their user network structures. Qualitative analysis demonstrates the descriptive power of \model{} through direct mapping of user and topic embeddings into a 2-D space. The explicit modeling  of dependency among user-generated data confirms the effectiveness of \model{}, as indicated by the model's predictive power in recovering missing links and modeling unseen documents. The learnt user representation also enables accurate content recommendation to users.


\subsection{Experiment Settings}

\textbf{Datasets.} We employed two large publicly available user-generated text datasets together with the associated user networks: 
1) \textbf{Yelp}, collected from Yelp dataset challenge \footnote{Yelp dataset challenge. \href{http://www.yelp.com/dataset\_challenge}{http://www.yelp.com/dataset\_challenge}}, consists of 187,737 Yelp restaurant reviews generated by 10,830 users. The Yelp dataset provides user friendship imported from their Facebook friend connections. Among the whole set of users, 10,194 of them have friends with an average of 10.65 friends per user. 
2) \textbf{StackOverflow}, collected from Stackoverflow.com \footnote{StackOverflow. \href{http://stackoverflow.com}{http://stackoverflow.com}}, consists of 244,360 forum discussion posts generated by 10,808 users. While there is no explicit network structure in StackOverflow dataset, we utilized the ``\textit{reply-to}'' information in the discussion threads to build a user network, because this relation suggests implicit social connections among users based on their expertise and technical topic interest. We ended up with 10,041 connected users, with an average of 5.55 connections per user. We selected 5,000 unigram and bigram text features based on Document Frequency (DF) in both datasets. We randomly split the data for 5-fold cross validation in all the reported experiments.

\textbf{Baselines.} We compared the proposed \model{} model against a rich set of user representation learning methods, including topic modeling based solutions, the network embedding methods, and models performing joint modeling of text and network. \noindent 1) \textbf{Latent Dirichlet Allocation (LDA)} \cite{blei2003latent} generates the topic distribution in documents across different users, and the user presentation is constructed by averaging the posterior topic proportion of documents associated with a user. \noindent 2) \textbf{Relational Topic Model (RTM)} \cite{chang2009relational} explicitly models the connection between two documents and we constructed a user-level network by concatenating all documents of one user in this baseline. \noindent 3) \textbf{Hidden Factors and Hidden Topics (HFT)} \cite{mcauley2013hidden} combines latent rating dimensions of users with latent review topics for user modeling. Users' ``\textit{upvote}'' toward a question is utilized as a proxy of rating in StackOverflow. \noindent 4) \textbf{Collaborative Topic Regression (CTR)} \cite{wang2011collaborative} combines collaborative filtering with topic modeling to explain the observed text content and ratings. \noindent 5) \textbf{DeepWalk (DW)} \cite{perozzi2014deepwalk} takes truncated random walks as input to learn social representations of vertices in the network.  \noindent 6) \textbf{Text-Associated DeepWalk (TADW)} \cite{yang2015network} further incorporates text content of vertices into network representation learning under the framework of joint matrix factorization.

\textbf{Parameter Settings.} We set the latent dimensions of user and topic embeddings to 10 in both JNET and baselines as larger dimension gives limited performance improvement but slows down all models considerably. As we tuned the topic size from 10 to 100, we found the learnt topics are most representative and meaningful at around 40 topics. Hence, we set topic number to 40 in the reported experiments. The maximum number of iteration in our EM algorithm is set to 100. Both the source codes and data are available online \footnote{JNET. \href{https://github.com/Linda-sunshine/JNET.}{https://github.com/Linda-sunshine/JNET.}}.

\subsection{The Learnt User Representations}
We first study the quality of the learnt user representations from \model{}. The learnt user embeddings are mapped to a 2-D space using the t-SNE algorithm and is visualized in Figure \ref{frame}. For illustration purpose, we simply assign each user to the topic that he/she is closest to, i.e., $\argmax_k(\phi_k\cdot u_i)$ and we mark users sharing the same interested topic with the same color. We also plot the most representative words of each topic learnt from \model{} (i.e., $\argmax_w p(w|\beta_z)$), with the same color of the corresponding set of users.

As we can find from the visualization of StackOverflow, users of similar interests are clearly clustered in the 2-D space, which indicates the descriptive power of our learnt user vectors. Meanwhile, we can easily identify the theme of each learnt topic, such as C++ (in light green circle), SQL (in dark purple circle) and java (in light blue circle). It is also interesting to find correlations among the users and topics by looking into their distances. The users in dark green are mainly interested in website development, thus are far away from the users who are interested in C++ (in light green). The users in orange care more about the network communication and they are overlapped with other clusters of users focusing on SQL (in dark purple) and C++ (in light green) as network communication is an important component among different programming languages. Similar observations can also be found on Yelp dataset.


\begin{figure}[t] 
\centering
\includegraphics[width=8.8cm]{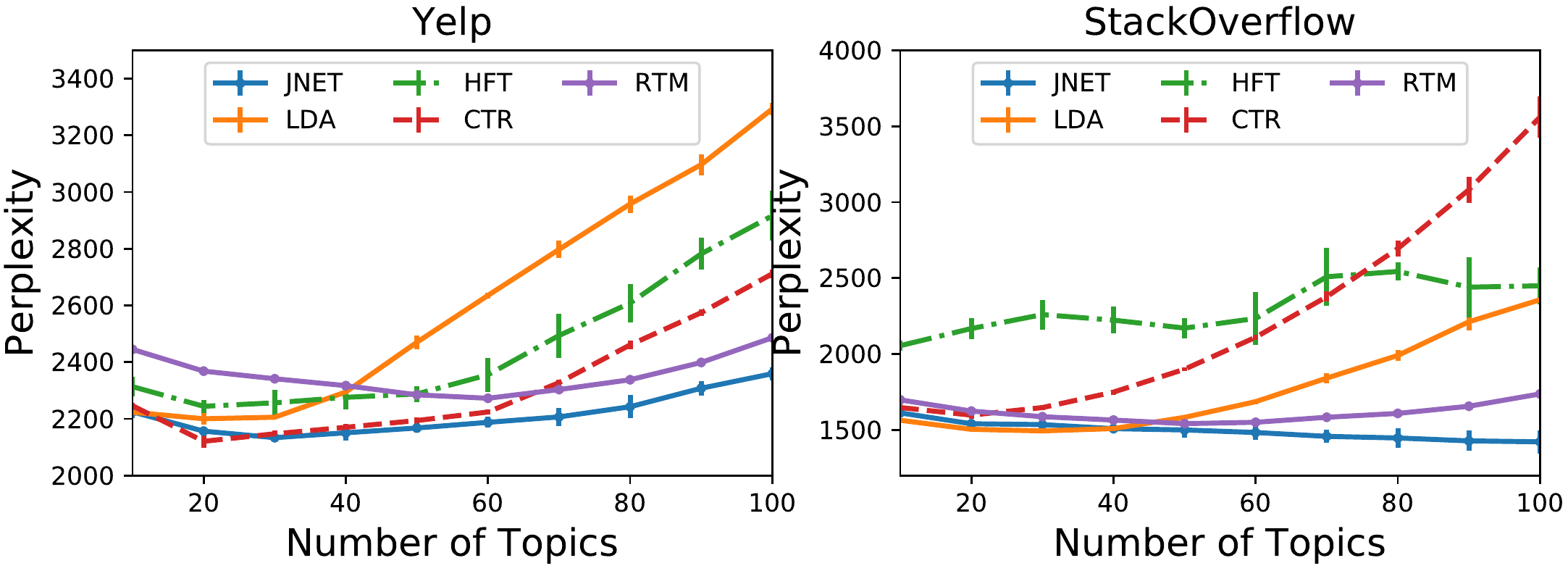} 
\vspace{-5mm}
\caption{Perplexity comparison on Yelp and StackOverflow.}
\label{perp1}
\vspace{-5mm}
\end{figure} 

\begin{figure*}[t]
\includegraphics[width=18cm]{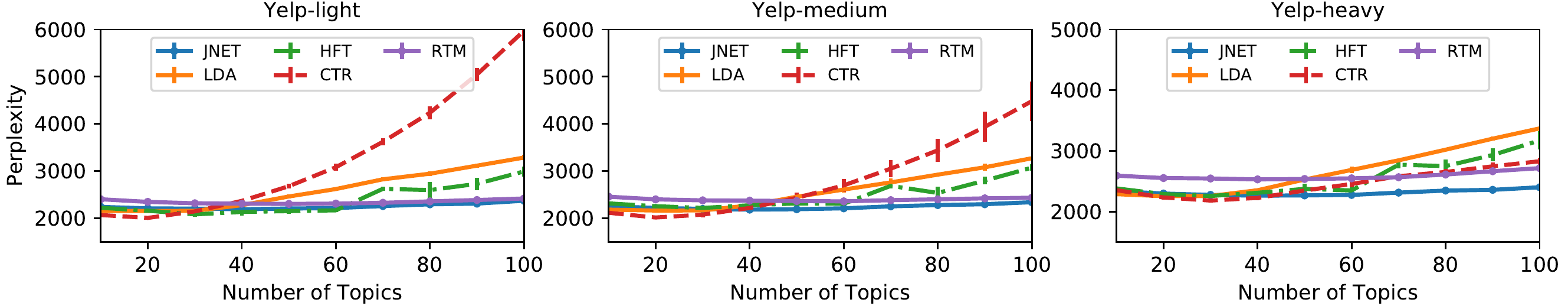}
\vspace{-5mm}
\caption{Comparison of perplexity in cold-start users on Yelp.}
\label{perp2}
\vspace{-3mm}
\end{figure*}

\subsection{Document Modeling}
In order to verify the predictive power of the proposed model, we first evaluated the generalization quality of \model{} on the document modeling task. We compared all the topic model based solutions by their \textit{perplexity} on a held-out test set.
Formally, the perplexity for a set of held-out documents is calculated as follows \cite{blei2003latent}:
\begin{equation*}
    perplexity(D_{test})=\exp{\big(-\frac{\sum_{d\in D_{test}}\log p(\bm{w}_{d})}{\sum_{d\in D_{test}}|d|}\big)}
\end{equation*}
where $p(\bm{w}_{d})$ is the likelihood of each held-out document given by a trained model. A lower perplexity indicates better generalization quality of a model.

\begin{figure*}[t]
\includegraphics[width=17.6cm]{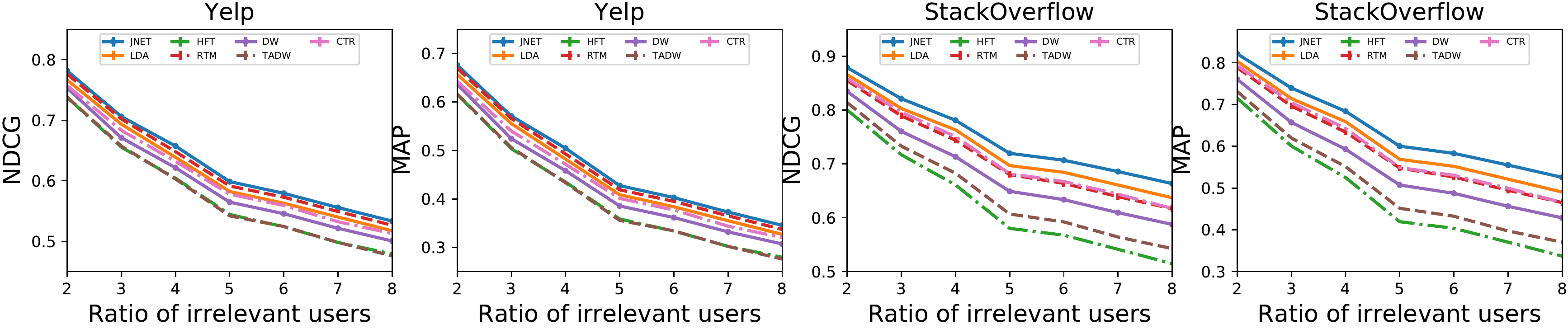}
\vspace{-2mm}
\caption{The performance comparison of link suggestion on Yelp and StackOverflow.}
\label{fig_link_pred}
\vspace{-2mm}
\end{figure*}

\begin{table*}[t]
\small
	\centering
	\caption{The performance comparison of link prediction for cold-start users on StackOvewrflow.}\label{tbl:batch}
	\vspace{-1mm}
	\begin{tabular}{|l|ccc|ccc|ccc|}
	\hline
	\multirow{3}{*}{Models} &\multicolumn{3}{c|}{Light}&\multicolumn{3}{c|}{Medium}&\multicolumn{3}{c|}{Heavy}\\
	& Ratio=2 & Ratio=4 & Ratio=6 & Ratio=2 & Ratio=4 & Ratio=6 & Ratio=2 & Ratio=4 & Ratio=6 \\
	&NDCG/MAP & NDCG/MAP & NDCG/MAP & NDCG/MAP & NDCG/MAP & NDCG/MAP & NDCG/MAP & NDCG/MAP & NDCG/MAP\\
	\hline
	LDA &0.786/0.648 &0.664/0.477 &0.632/0.431 &0.774/0.597 &0.677/0.451 &0.612/0.364 &0.818/0.581 &0.745/0.443 &0.697/0.366\\
	HFT &0.666/0.493 &0.543/0.333 &0.483/0.259 &0.671/0.461 &0.562/0.313 &0.492/0.226 &0.682/0.389 &0.591/0.250 &0.532/0.179\\
	RTM &0.777/0.642 &0.688/0.514 &0.627/0.433 &0.801/0.638 &0.709/0.495 &0.654/0.419 &0.837/0.624 &0.760/0.481 &0.711/0.399\\
	TADW &0.695/0.525 &0.583/0.373 &0.515/0.291 &0.696/0.481 &0.591/0.336 &0.532/0.263 &0.739/0.448 &0.639/0.298 &0.587/0.229\\
	\model&$\textbf{0.794/0.664}$ &$\textbf{0.697/0.534}$ &$\textbf{0.643/0.453}$ &$\textbf{0.812/0.649}$ &$\textbf{0.724/0.511}$ &$\textbf{0.663/0.425}$ &$\textbf{0.842/0.626}$ &$\textbf{0.763/0.483}$ &$\textbf{0.713/0.399}$\\
	\hline
	\end{tabular}
\vspace{-2mm}
\end{table*}

Figure \ref{perp1} reports the mean and variance of the perplexity for each model with 5-fold cross validation over different topic sizes. \model{} achieved the best predictive power on the hold-out dataset, especially when an appropriate topic size is assigned. RTM achieved comparative performance as it utilizes the connectivity information among users, but it is limited by not being able to capture the variance within each user's different documents. The other baselines do not explicitly model network data, i.e., LDA, HFT and CTR, and therefore suffer in their performance. 

A good joint modeling of network structure and text content should complement each other to facilitate a more effective user representation learning. Hence, we expect a good model to learn reasonable representations on users lacking text information, a.k.a., cold-start users, by utilizing network structure. We randomly selected 200 users and held out all their text content for testing. Regarding to the number of social connections each testing user has in training data, we further consider three different sets of users, and name them as light, medium and heavy users, to give a finer analysis with respect to the degree of connectivity in cold-start setting. The threshold for categorizing different sets of users is based on the statistics of each dataset; and each group contains 200 users. In particular, we selected 5 and 20 as the connectivity threshold for Yelp, 5 and 15 as the threshold for StackOverflow respectively. That is, in Yelp, light users have fewer than 5 friends, medium users have more than 5 friends while fewer than 20 friends and heavy users have more than 20 friends. We compared \model{} against four baselines, i.e., LDA, HFT, RTM, CTR for evaluation purpose. We reported the perplexity on the held-out test documents regarding to the three sets of users, in Figure~\ref{perp2}. 

As we can observe in Figure~\ref{perp2}, \model{} performed consistently better on the testing documents for the three different sets of unseen users on Yelp dataset, which indicates the advantage of utilizing network information in addressing cold-start content prediction issue. The benefit of network is further verified across different sets of users as heavily connected users can achieve better performance improvement compared with text only user representation model, i.e., LDA. Similar conclusion is obtained for StackOverflow dataset, while we neglect it due to the space limit.

\subsection{Link Prediction}
The predictive power of \model{} is not only reflected in unseen documents, but also in missing links. In the task of link prediction, the key component is to infer the similarity between users. 
We split the observed social connections into 5 folds. Each time, we held out one fold of edges for testing and utilized the rest for model training, together with users' text content. In order to construct a valid set of ranking candidates for each testing user, we randomly injected irrelevant users (non-friends) for evaluation purpose. And the number of irrelevant users is proportional to the number of connections a testing user has, i.e., $t\times \text{number of social connections}$. 
We rank users based on the cosine similarity between their embedding vectors. Normalized discounted cumulative gain (NDCG) and mean average precision (MAP) are used to measure the quality of ranking. We started with the ratio between irrelevant users and relevant users being $t=2$ and increased the ratio to $t=8$ to make the task more challenging to further verify the effectiveness of the learnt user representations. 

To compare the prediction performance, we tested five baselines, i.e., LDA, HFT, RTM, DW and TADW. 
We reported the NDCG and MAP for the two datasets in Figure~\ref{fig_link_pred}. It is clear \model{} achieved encouraging performance on both datasets, which indicates effective user representations are learnt to recover network structure. In Yelp dataset, network-only solutions, i.e., DW, and text-only solutions, i.e., LDA and HFT, cannot take the full advantage of both modalities of user-generated data to capture user intents, while RTM achieved descent performance due to the integration of content and network modeling. Since the way of constructing network in StackOverflow is more content oriented, the performance of link prediction on StackOverflow prefers the text based solutions, which explains the comparable performance of LDA. Though TADW utilizes both modalities for user modeling, it fails to capture the dependency between them, leading to the poor performance on this task.

In practice, link prediction for unseen users is especially useful. For example, friend recommendation for new users in a system: they have very few or no friends, while they may associate with rich text content. This is also known as ``cold-start'' link prediction. Network-only solutions will suffer from the lack of information in such users. However, a joint model can overcome this limitation by utilizing user-generated text content to learn representative user vectors, thus to provide helpful link prediction results.

In order to study the models' predictive power in the cold-start setting, we randomly sampled three sets of users, regarding to the number of documents each user has, and name them as light, medium and heavy users accordingly. Each set of users consists of 200 users, and we selected 10 and 50 as the threshold for Yelp, 15 and 50 as the threshold for StackOverflow respectively. For example, in StackOverflow, light users have fewer than 15 posts, medium users have more than 15 but fewer than 50 posts, and heavy users have more than 50 posts. We compared \model{} against four baselines, i.e., LDA, HFT, RTM and TADW for evaluation purpose. Because DW cannot learn representations for users without any network information, it is excluded in this experiment. We also randomly injected irrelevant users as introduced before for evaluation and we varied the ratio to change the difficulty of the task. We reported the NDCG and MAP performance on the three sets of users in Stackoverflow dataset with three different ratios, i.e., 2, 4 and 6, in Table~\ref{tbl:batch}, respectively. 

\model{} achieved consistently favorable performance in cold-start users, as accurate proximity between users is properly identified with its user representations learnt from text data. Comparing across user groups, better performance is achieved for users with more text documents. Similar results were obtained on Yelp dataset as well, but omitted due to space limit.


\subsection{Expert Recommendation}
In the sampled StackOverflow dataset, the average number of answers for questions is as low as 1.14, which indicates the difficulty for getting an expert to answer the question. If the system can suggest the right user to answer the posted questions, e.g., push the question to the selected user, more questions would be answered more quickly and accurately. We conjecture the learnt topic distribution of each question in StackOverflow, together with the identified user representation, can facilitate the task of expert recommendation for question answering. The task can be further decomposed into two components: whether the question falls into a user's skill set; and whether the user who asked the question shares similar interests with the potential candidate experts. With the learnt topic embeddings $\Phi$ and each user's embedding $u_i$, each user's interest over topics can be characterized as a mapping from the topic embeddings to the user's embedding, i.e., $\Phi\cdot u_i$. Together with the learnt topic distribution of each question, we can estimate the proximity between a question and a user's expertise to score the alignment between them. In the meanwhile, the closeness between users can be simply measured by the distance of their corresponding embedding vectors. As a result, the task can be formalized as finding the user that achieves the highest relatedness with the given question, where we define the relatedness as follows:
\begin{equation}
\label{eq_matching}
\text{score}=\alpha \cdot \text{cosine}(u_i\cdot \Phi, \theta_{id}) + (1-\alpha)\cdot \text{cosine}(u_i, u_j)
\end{equation}

\begin{figure}[t]
\includegraphics[width=8.6cm]{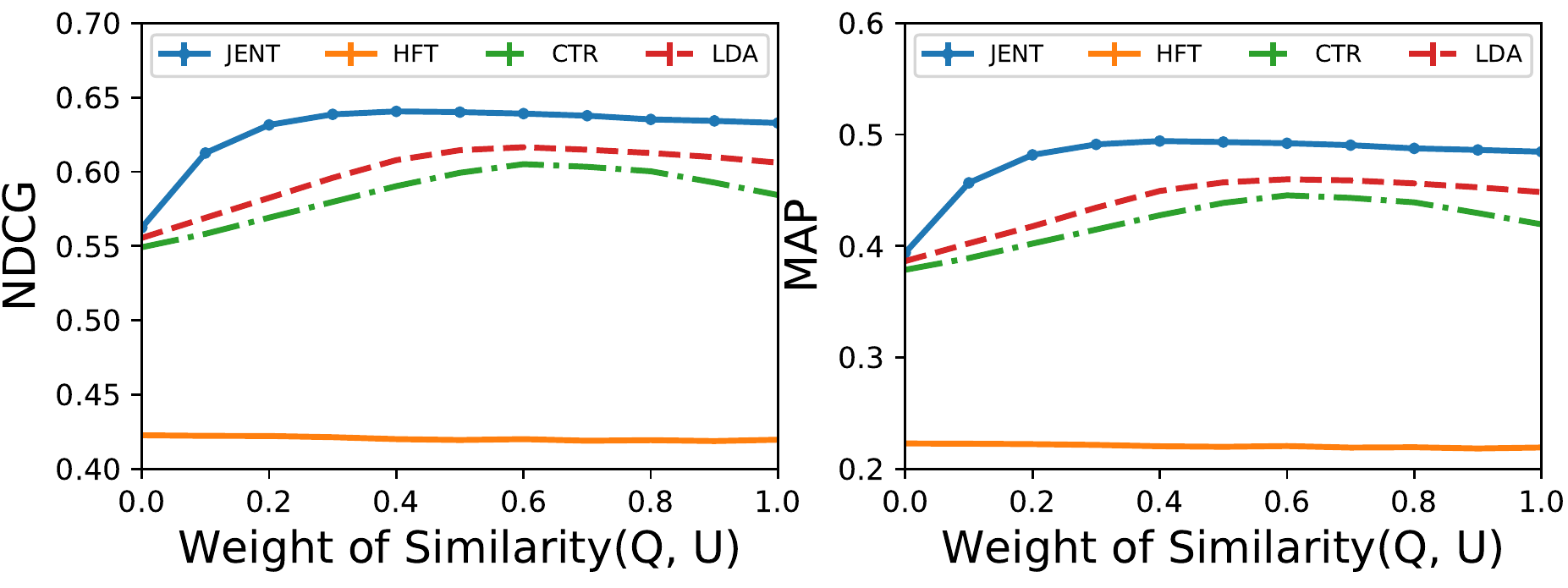}
\vspace{-4mm}
\caption{Expert recommendation on StackOverflow.}
\label{fig_ansrec}
\vspace{-4mm}
\end{figure}

Due to the limited number of answers for each question in our dataset, we selected 1,816 questions with more than 2 answers for the experiment. Besides the users that answered the given question, we also incorporated irrelevant users for each question for evaluation purpose. And the number of irrelevant users is 10 times of the number of answers. We compared against the learnt topic distributions of questions and user representations from LDA, HFT and CTR as we cannot get the topic distribution of each question from the other baselines.
As we tune the weight between the two components in Eq \eqref{eq_matching}, we plot the corresponding NDCG and MAP in Figure ~\ref{fig_ansrec}.

\model{} achieved very promising performance in this recommendation task, as it explicitly models a user's expertise and a given question in the topic space. The estimated similarities between user-user and user-content accurately align the question to the right user. The baseline models can only capture the similarity between questions and users based on their topical similarity, which is insufficient in this task. Interestingly, as we gradually increased the weight of question-content similarity from 0 to 1, \model{}'s performance peaked, which indicates the relative importance between user-user and user-content similarities for this specific problem.
\section{Conclusion and Future Work}
In the paper, we studied user representation learning by explicitly modeling the structural dependency among different modalities of user-generated data. We proposed a complete generative model named \model{} to integrate user representation learning with content modeling and social network modeling. The learnt user representations are interpretable and predictive, indicated by the performance improvement in many important tasks such as link prediction and expert recommendation. 

Several areas are left open for our future explorations. The current model focuses on the first-order proximity among users in network modeling, while higher-order proximity can be explored to better capture the network connectivity. Also, temporal information of the text content and connections are not considered in the current model. By properly utilizing the temporal information, we would be able to learn the dynamics of user representations, together with the evolution of topics and social network.

\section{Acknowledgments}
We thank the anonymous reviewers for their insightful comments. This paper is based upon work supported by the National Science Foundation under grant IIS-1553568, IIS-1718216.

\bibliographystyle{ACM-Reference-Format}
\bibliography{sigproc}  
%
%

\appendix
\newpage
\section{Variational Inference}
\label{app_inf}
In this section, we provide the detailed derivation of the likelihood lower bound in Eq \eqref{eq_elbo}. 

Recall that we begin by postulating a factorized distribution:
\begin{align*}
q(\Phi, U, \Delta, \Theta, Z)=& \prod\nolimits_{k=1}^Kq(\phi_k) \prod\nolimits_{i=1}^Uq(u_i)\\
&\Big[\prod\nolimits_{j=1,j\neq i}^Uq(\delta_{ij})\prod\nolimits_{d=1}^Dq(\theta_{id})\prod\nolimits_{n=1}^Nq(z_{idn})\Big]
\end{align*}
where the factors have the following parametric forms:
\begin{align}
&q(\phi_k)=\mathcal{N}(\phi_k|\mu^{(\phi_k)},\Sigma^{(\phi_k)}), q(u_i)=\mathcal{N}(u_i|\mu^{(u_i)},\nonumber
\Sigma^{(u_i)}),\nonumber\\&q(\delta_{ij})=\mathcal{N}(\delta_{ij}|\mu^{(\delta_{ij})}, {\sigma^{(\delta_{ij})}}^2),\nonumber
q(\theta_{id})=\mathcal{N}(\theta_{id}|\mu^{(\theta_{id})},\Sigma^{(\theta_{id})}),\\&q(z_{idn})=\text{Mult}(z_{idn}|\eta_{idn})\nonumber
\end{align}

The log likelihood of observed user behaviors, e.g., posted texts and connected social relations, is 
bounded by a lower bound using Jensen's inequality: 
\begin{align*}
&\log p(\bm w, \bm e |\alpha, \beta, \gamma, \tau)
\\
& \geq \mathbb{E}_{q}[\log p(U, \Theta, Z, \Phi, \Delta, \bm w, \bm e|\alpha, \beta, \gamma, \tau)] -\mathbb{E}_{q}[\log q(U, \Theta, Z, \Phi, \Delta)]\\
&=\mathbb{E}_{q}[\log p(\Phi|\alpha)] + \mathbb{E}_{q}[\log p(U|\gamma)] + \mathbb{E}_{q}[\log p(\Delta|U)] + \mathbb{E}_{q}[\log p(\bm{e}|\Delta)]\\
& + \mathbb{E}_{q}[\log p(\Theta|U, \Phi, \tau)] + \mathbb{E}_{q}[\log p(Z|\Theta)] + \mathbb{E}_{q}[\log p(\bm{w}|Z, \beta)]\\
&- \mathbb{E}_{q}[\log q(\Phi, U, \Delta, \Theta, Z)]
\end{align*}

Thanks to the conjugate priors introduced on $U$ and $\Phi$, the expectations related to these latent variables are straightforward. However, the calculations of $\mathbb{E}_{q}[\log p(\bm{e}|\Delta)]$ and $\mathbb{E}_{q}[\log p(Z|\Theta)]$ are difficult due to no conjugate prior for logistic Normal distribution. We will first provide details of these two nontrivial expectations, and then list the other terms for reproducibility.

\noindent\textbf{$\bullet$ Compute $\mathbb{E}_{q}[\log p(\bm{e}|\Delta)]$.} The nonconjugacy of logistic normal leads to difficulty in computing the expected probability of edge assignment between $u_{i}$ and $u_{j}$:
\begin{align*}
\mathbb{E}_{q} [\log p(e_{ij}|\delta_{ij})]=\mathbb{E}_{q} [e_{ij}\delta_{ij}]-\mathbb{E}_{q}[\log(1+\exp(\delta_{ij}))]
\end{align*}

We utilize the inequality of logarithm $\log x\leq x-1$, and set $x=\varepsilon^{-1}(1+\exp(\delta_{ij}))$, to approximate the second term:
\begin{align*}
& \log(1+\exp(\delta_{ij})) \leq \varepsilon^{-1}(1+\exp(\delta_{ij}))-1+\log\epsilon \nonumber
\end{align*}
Thus, the corresponding expectation is as follows,
\begin{align*}
\label{epsilon}
\mathbb{E}_{q} [\log (1+\exp(\delta_{ij}))]&\leq \varepsilon^{-1}\mathbb{E}_{q}[1+\exp(\delta_{ij})]-1+\log \varepsilon
\end{align*}
where the expectation is mean of log normal:  $$\mathbb{E}_{q}[1+\exp(\delta_{ij})]=1+\exp(\mu^{(\delta_{ij})}+\frac{1}{2}{\sigma^{(\delta_{ij})}}^2)$$
Put them together, we get the expectation as follows:
\begin{align*}
&\mathbb{E}_{q}[\log p(e_{ij}|\delta_{ij})]\\
&\ge e_{ij}\mu^{(\delta_{ij})}-\varepsilon^{-1}(1+\exp(\mu^{(\delta_{ij})}+\frac{1}{2}{\sigma^{(\delta_{ij})}}^2))+1-\log \varepsilon
\end{align*}
where a new variational parameter $\varepsilon$ is introduced, and we set $\varepsilon=1+exp(\delta_{ij})$ to approach the equality.

\noindent\textbf{$\bullet$ Compute $\mathbb{E}_{q}[\log p(Z|\Theta)]$.} The nonconjugacy of logistic normal also exists in computing the expectation of topic assignment for each word of $u_{i}$'s $d$-th document:
\begin{align*}
\mathbb{E}_{q} [\log p(z_{idn}|\theta_{id})]=\mathbb{E}_{q} [\theta_{id}^\mt z_{idn}]-\mathbb{E}_{q} [\log(\sum\nolimits_{k=1}^K\exp(\theta_{idk}))]
\end{align*}

We again utilize the equality of logarithm $\log x \le x-1$, and set $x=\zeta^{-1}\sum_{k=1}^K\exp(\theta_{idk})$ to compute the second term:
\begin{equation*}
\log \sum\nolimits_{k=1}^K\exp(\theta_{idk}) \leq \zeta^{-1}\sum\nolimits_{k=1}^K\exp(\theta_{idk})-1+\log \zeta
\end{equation*}
Thus the second term is calculated as:
\begin{align}
\mathbb{E}_{q} [\log (\sum\nolimits_{k=1}^K\exp(\theta_{idk}))] \leq \zeta^{-1}(\sum\nolimits_{k=1}^K\mathbb{E}_{q}[\exp(\theta_{idk})])-1+\log \zeta\nonumber
\end{align}
where the expectation is mean of log normal distribution: $$\mathbb{E}_{q}[\exp(\theta_{idk})]=\exp({\mu_k}^{(\theta_{id})}+\frac{1}{2}\Sigma_{kk}^{(\theta_{id})})$$
Putting them together, we get the expectation as follows:
\begin{align*}
&\mathbb{E}_{q} [\log p(z_{idn}|\theta_{id})]\\
&\ge{\mu^{(\theta_{id})}}^\mt  \eta_{idn}-\zeta^{-1}\sum\nolimits_{k=1}^K\exp({\mu_k}^{(\theta_{id})}+\frac{1}{2}\Sigma_{kk}^{(\theta_{id})})+1-\log \zeta
\end{align*}
where another variational parameter $\zeta$ is introduced, and we set $\zeta=\sum_{k=1}^K\exp(\theta_{idk})$ to approach the equality. 

\noindent\textbf{$\bullet$ Compute $\mathbb{E}_{q}[\log p(\Phi|\alpha)]$.} Topic embedding follows Gaussian distributions for $p$ and $q$, and the corresponding expectation is:
\begin{align*}
\mathbb{E}_{q}[\log p(\phi_k|\alpha)]
\propto \frac{M}{2}\log \alpha - \frac{\alpha}{2}[\sum\nolimits_{m=1}^M\Sigma_{mm}^{(\phi_k)}+{\mu^{(\phi_k)}}^\mt \mu^{(\phi_k)}]
\end{align*}

\noindent\textbf{$\bullet$ Compute $\mathbb{E}_{q}[\log p(U|\gamma)]$.} User embedding also follows Gaussian distributions, thus:
\begin{equation*}
\mathbb{E}_{q}[\log p(u_i|\gamma)]\propto \frac{M}{2}\log \gamma - \frac{\gamma}{2}[\sum\nolimits_{m=1}^M\Sigma_{mm}^{(u_i)}+{\mu^{(u_i)}}^\mt \mu^{(u_i)}]
\end{equation*}

\noindent\textbf{$\bullet$ Compute $\mathbb{E}_{q}[\log p(\Delta|U)]$.} The affinity $\delta_{ij}$ between a pair of users $u_{i}$ and $u_{j}$ follows Gaussian Distributions. Thus, the corresponding expectation can be written as follows:
\begin{align*}
&\mathbb{E}_{q} [\log p(\delta_{ij}|u_i, u_j)]\\
&\propto -\log \xi-\frac{1}{2\xi^2}\mathbb{E}_{q} [(\delta_{ij}-u_i^\mt u_j)^2]\nonumber
\\&=-\log \xi-\frac{1}{2\xi^2}\mathbb{E}_{q}[{\delta_{ij}^2} ]+ \frac{1}{\xi^2}\mathbb{E}_{q}[u_i^\mt u_j]\mathbb{E}_{q}[\delta_{ij}]-\frac{1}{2\xi^2}\mathbb{E}_{q}[(u_i^\mt u_j)^2]
\end{align*}
where the expectations of Gaussian distributions for $\delta$ and $u$ can be directly written, thus we get:
\begin{align*}
&\mathbb{E}_{q} [\log p(\delta_{ij}|u_i, u_j)]\\
&\propto -\log \xi-\frac{1}{2\xi^2}({\mu^{(\delta_{ij})}}^2+{\sigma^{(\delta_{ij})}}^2)
+ \frac{\mu^{(\delta_{ij})}}{\xi^2}{\mu^{(u_i)}}^\mt\mu^{(u_j)}\\
&-\frac{1}{2\xi^2}(\Sigma^{(u_i)}+\mu^{(u_i)}{\mu^{(u_i)}}^\mt)^\mt(\Sigma^{(u_j)}+\mu^{(u_j)}{\mu^{(u_j)}}^\mt)
\end{align*}

\noindent\textbf{$\bullet$ Compute $\mathbb{E}_{q}[\log p(\Theta|U, \Phi, \tau)]$.} The topic proportion of each user's document follows Gaussian distribution. The corresponding expectation can be written as:
\begin{align*}
&\mathbb{E}_{q} [\log p(\theta_{id}|u_i, \Phi, \tau)]\\
&\propto{\frac{K}{2}}\log \tau-\frac{\tau}{2}\{\mathbb{E}_{q} [\theta_{id}^\mt \theta_{id}]-\mathbb{E}_{q} [\theta_{id}^\mt \Phi u_i]-\mathbb{E}_{q} [u_i^\mt \Phi^\mt \theta_{id}]+\mathbb{E}_{q} [u_i^\mt \Phi^\mt \Phi u_i]\}
\end{align*}
where $\Phi$ and $u$ also follows Gaussian and the calculation is straightforward, thus we get:
\begin{align*}
&\mathbb{E}_{q} [\log p(\theta_{id}|u_i, \Phi, \tau)]\\
&\propto {\frac{K}{2}}\log \tau-\frac{\tau}{2}[\sum\nolimits_{k=1}^K\Sigma_{kk}^{(\theta_{id})}+{\mu^{(\theta_{id})}}^\mt \mu^{(\theta_{id})}]+\tau \sum\nolimits_{k=1}^K\mu_k^{(\theta_{id})}{\mu^{(\phi_k)}}^\mt\mu^{(u_i)}
\\
&-\frac{\tau}{2}\sum\nolimits_{k=1}^K(\Sigma^{(u_i)}+\mu^{(u_i)}{\mu^{(u_i)}}^\mt)^\mt(\Sigma^{(\phi_k)}+\mu^{(\phi_k)}{\mu^{(\phi_k)}}^\mt)
\end{align*}

\noindent\textbf{$\bullet$ Compute $\mathbb{E}_{q}[\log p(\bm{w}|Z, \beta)]$.}
The expectation of word assignment is given by:
\begin{align*}
\mathbb{E}_{q}[\log p(w_{idn}|z_{idn}, \beta)]&=\sum\nolimits_{k=1}^K\sum\nolimits_{v=1}^V\mathbb{E}_{q}[z_{idn}^k w_{idn}^v\log \beta_{kv}]\\
&=\sum\nolimits_{k=1}^K\sum\nolimits_{v=1}^Vw_{idn}^v \eta_{id,k} \log \beta_{kv}
\end{align*}

\section{Parameter Estimation}
\label{para_est}
By taking the gradient of $\mathcal{L}(q)$ in Eq~\eqref{eq_elbo} with respect to the variance of user affinity $\xi$, and set it to 0, we get the closed form estimation as follows:
\begin{align*}
\xi^2 = & [U(U-1)]^{-1}\sum\nolimits_{i=1}^U\sum\nolimits_{j\neq i}^U\Big[{\mu^{(\delta_{ij})}}^2+{\sigma^{(\delta_{ij})}}^2-2\mu^{(\delta_{ij})}{\mu^{(u_i)}}^\mt\mu^{(u_j)}
\\&+(\Sigma^{(u_i)}+\mu^{(u_i)}{\mu^{(u_i)}}^\mt)^\mt(\Sigma^{(u_j)}+\mu^{(u_j)}{\mu^{(u_j)}}^\mt)\Big]
\end{align*}
Similarly, the closed form estimation for the variance of document topic proportion $\tau$ is given by:
\begin{align*}
&\tau^{-1} = (KD)^{-1}\sum\nolimits_{i=1}^U\sum\nolimits_{d=1}^{D_i}\Big[(\sum\nolimits_{k=1}^K\Sigma_{kk}^{(\theta_{id})}+{\mu^{(\theta_{id})}}^\mt \mu^{(\theta_{id})})\\
&-\sum\nolimits_{k=1}^K(2\mu_{k}^{(\theta_{id})}{\mu^{(\phi_k)}}^\mt\mu^{(u_i)}-(\Sigma^{(u_i)}+\mu^{(u_i)}{\mu^{(u_i)}}^\mt)^\mt(\Sigma^{(\phi_k)}+\mu^{(\phi_k)}{\mu^{(\phi_k)}}^\mt))]
\end{align*}

\end{document}